\documentclass[aps,pre,floats,floatfix,10pt,onecolumn,superscriptaddress,longbibliography,notitlepage,showkeys]{revtex4-1}
\usepackage{bm}
\usepackage{mathrsfs}
\usepackage{latexsym,amsmath}
\usepackage{amsbsy}
\usepackage{amssymb}
\usepackage{bm}
\usepackage{hyperref}
\usepackage[pdftex]{graphicx}
\usepackage[usenames]{color}
\usepackage[normalem]{ulem}
\usepackage{microtype}
\usepackage{epstopdf}
\begin{document}
\title{Detecting the \textit{Escherichia coli} metabolic backbone}

\author{Oriol G{\"u}ell}\email{oguellri8@alumnes.ub.edu}\affiliation{Departament de Qu\'{i}mica F\'{i}sica, Universitat de Barcelona, Mart\'{i} i Franqu\`{e}s 1, 08028 Barcelona, Spain}
\author{Francesc Sagu{\'e}s}\affiliation{Departament de Qu\'{i}mica F\'{i}sica, Universitat de Barcelona, Mart\'{i} i Franqu\`{e}s 1, 08028 Barcelona, Spain}
\author{M. {\'A}ngeles Serrano}\affiliation{Departament de F\'{i}sica Fonamental, Universitat de Barcelona, Mart\'{i} i Franqu\`{e}s 1, 08028 Barcelona, Spain}\affiliation{Instituci\'o Catalana de Recerca i Estudis Avan\c{c}ats (ICREA), Passeig Llu\'is Companys 23, E-08010 Barcelona, Spain}

\begin{abstract}
The heterogeneity of reaction fluxes present in a metabolic network within a single flux state can be exploited to construct the so-called backbone as a reduced version of metabolism. The backbone maintains all significant fluxes producing or consuming metabolites while displaying a substantially decreased number of interconnections and, hence, it becomes a useful tool to extract primary metabolic routes. Here, we disclose the metabolic backbone of \textit{Escherichia coli} using the computationally predicted fluxes which maximize the growth rate in glucose minimal medium, and we compare it with the backbone of \textit{Mycoplasma pneumoniae}, a much simpler organism. We find that the central core in both backbones is mainly composed of reactions in ancient pathways, still playing at present a key role in energy metabolism. In \textit{E. coli}, the analysis of the backbone reveals that the synthesis of nucleotides and the metabolism of lipids form smaller cores which rely critically on energy metabolism; but not conversely. At the same time, an analysis of the dependence of this backbone on media composition leads to the identification of pathways sensitive to environmental changes. The metabolic backbone of an organism is thus useful to trace simultaneously both its evolution and adaptation fingerprints.
\end{abstract}


\maketitle

\section{Introduction}
High-quality genome-scale metabolic reconstructions are composed of thousands of reactions and metabolites \cite{Feist:2006a,Feist:2007a,Schellenberger:2010a,Orth:2011a}.  Due to their complexity, the analysis of these metabolic reconstructions requires computational approaches like constraint-based optimization techniques \cite{Orth:2010a,Gudmundsson:2010a} and methodological frameworks like complex network science \cite{Barrat:2008,Newman:2010} to elucidate features of their functional organization and pathway structure. Some of the tools used with this purpose are Elementary Flux Modes \cite{Schuster:1994}, Extreme Pathways \cite{Price:2003a}, Minimal Metabolic Behaviors \cite{Larhlimi:2009a}, and Minimal Pathway Structures \cite{Bordbar:2014}, which are based on finding feasible subnetworks to relate them with definite functions. A different approach in this endeavor is provided by the concept of backbone. Backbones maintain significant information while displaying a substantially decreased number of interconnections and, hence, can provide accurate but reduced versions of the whole system. In this direction, the work by Almaas \textit{et al.} \cite{Almaas:2004a} introduced a filtering technique that selects the reaction that dominates the production or consumption of each metabolite. This method is able to segregate classical pathways, but the selected high-flux subgraphs present a linear structure with very little interconnectivity and so they necessarily lack the characteristic complex features of real metabolic networks \cite{Ma:2003b,Barabasi:2004a,Serrano:2012a}. 

Filtering approaches have also interested researchers working on networks in a more general context. A filtering method for weighted networks based on the disparity measure \cite{HHIHerfindahl:1959,HHIHirschman:1964} was developed in Ref. \cite{Serrano:2009a}. This approach exploits the heterogeneity present in the intensity of interactions (weights) in real networks, both at the global and local levels \cite{Vespignani:2004a}, to extract the dominant set of connections for each element. Typically, the obtained disparity backbones preserve almost all nodes in the initial network and a large fraction of the total weight, while reducing considerably the number of links that pass the filter. At the same time, disparity backbones preserve the heterogeneity of the degree distribution, the level of clustering, and the bow-tie structure \cite{Broder:2000}, and other complex features of the original networks \cite{Serrano:2009a}. 

In this work, we use Flux Balance Analysis (FBA) \cite{Orth:2010a} maximizing the growth rate to determine reaction fluxes in the \textit{i}JO1366 version of the metabolic network of \textit{Escherichia coli} K-12 MG1655 \cite{Orth:2011a}, and afterwards we use the disparity filter \cite{Serrano:2009a} to extract its backbone. In contrast to the filtered linear structures from Ref. \cite{Almaas:2004a}, metabolic backbones obtained using the disparity filter conserve not only high-flux reactions but also many low-flux reactions --provided that they are significant for the production or consumption of a certain metabolite--, such that the complexity of the original networks is preserved in the backbone. We investigate the obtained backbone in glucose minimal medium for fingerprints of evolution and environmental adaptation finding that its central core is mainly composed of evolutionary conserved reactions in energy metabolism whose fluxes still retain at present a key role in the evolved organisms. This feature is also observed in the central core reactions of the backbone of the \textit{i}JW145 metabolic network of \textit{Mycoplasma pneumoniae} \cite{Wodke:2013a}, a much simpler organism. In \textit{E. coli}, the analysis of the backbone reveals that the synthesis of nucleotides and the metabolism of lipids form smaller cores which rely critically on energy metabolism; but not conversely. We also study how the structure of the backbone in \textit{E. coli} depends on the composition of the minimal media, which allows us to identify pathways that are more sensitive to environmental changes and nutrient availability.

\section{Results}
We use FBA to compute the fluxes of the reactions in a metabolic network which maximize the growth rate of the organism (see Methods, section \ref{SA}). The fluxes are treated as weights by the disparity filter (see Methods, section \ref{SB}). The large-scale connectivity structure of the obtained backbone is analyzed in terms of connected components (see Methods, section \ref{SC}). Additional media are considered to analyze the environmental sensitivity of the backbone composition (see Methods, section \ref{SD}).

\subsection{Identification of the disparity backbone of a metabolic network}
\begin{figure*}
 \centering
 \includegraphics[width=0.8\textwidth]{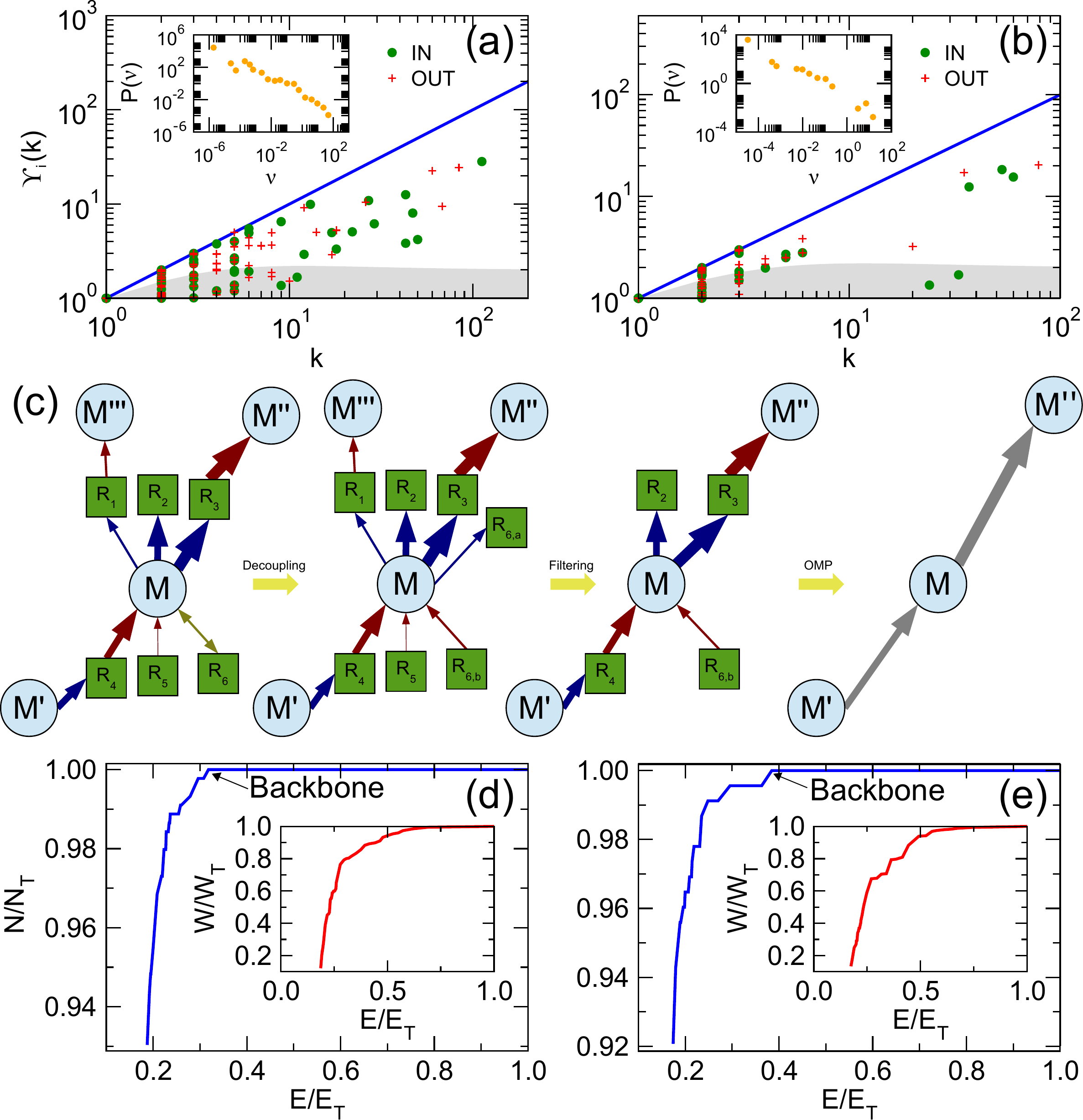}
 \caption{\textbf{Disparity filter and measures of the heterogeneity of reaction fluxes in \textit{E. coli} and \textit{M. pneumoniae}.} (a) Disparity measure as a function of incoming and outgoing degrees ($k$) in \textit{E. coli}. The shadowed area corresponds to the average plus 2 standard deviations given by the null hypothesis, meaning that points which lie outside this area are can be considered heterogeneous \cite{Serrano:2009a}.  It is important to notice that there may exist overlap between points for each value of $k$. Inset: global distribution of fluxes of \textit{E. coli}. (b) The same as in (b) for \textit{M. pneumoniae}. (c) Scheme of the filtering method. Blue nodes are metabolites and green squares denote reactions. Incoming connections to metabolites are represented by red arrows, outgoing connections with blue arrows, and bidirectional connections with dark yellow arrows. The width of the arrows is proportional to the value of the fluxes, \textit{i.e.}, the larger the flux, the wider the arrow. OMP denotes One-Mode Projection. (d) Fraction of metabolites as a function of the fraction of links in the filtered networks of \textit{E. coli}. Inset: remaining fraction of flux as a function of the fraction of links. Notice that the curves come from a parametric representation of the different fractions as a function of $\alpha$. (e) The same as in (d) for \textit{M. pneumoniae}.}
 \label{fig:1}
\end{figure*}

An important feature of the flux solutions obtained using FBA is that they capture the heterogeneity of the flux distribution within a single flux state \cite{Almaas:2004a}. The probability distribution function of the obtained FBA fluxes, insets of Figs. \ref{fig:1}a and \ref{fig:1}b, is characterized by a broad scale distribution of values. We disregarded zero-flux reactions, such that the set of active reactions and metabolites is markedly reduced as compared to their number when all reactions in the genome-scale reconstruction are considered, from 411 active reactions and 445 metabolites to 404 reactions and 445 metabolites in \textit{E. coli}, and from 163 active reactions and 227 metabolites to 162 reactions and 227 metabolites in \textit{M. pneumoniae}. Notice that all metabolites and nearly all reactions are conserved for both organisms, meaning that we do not lose information about the constituents of both metabolic networks. Heterogeneity is also present at the local level in the fluxes that produce and consume a given metabolite. We calculate the disparity measure for every metabolite $i$ participating in $k$ producing or consuming reactions \cite{Almaas:2004a,Serrano:2009a} as $\Upsilon_i(k) =  k \sum_{j=1}^k p_{ij}^2$, where $p_{ij}$ is the flux of reaction $j$ normalized by the total flux consuming or producing metabolite $i$, $p_{ij}=\nu_{ij}/\sum_{j=1}^k  \nu_{ij}$, (see Methods, section \ref{SB}). We treat separately producing and consuming reactions. Figs. \ref{fig:1}a and \ref{fig:1}b display the disparity values for all metabolites as a function of their number of producing and consuming reactions (incoming and outgoing degree) in \textit{E. coli} and in \textit{M. pneumoniae}, respectively. The shadowed areas correspond to disparity values compatible with the null hypothesis that the total incoming and outgoing flux is uniformly distributed at random among the incoming and outgoing reactions in which the metabolite participates. The null hypothesis helps to discount local heterogeneities produced by random fluctuations (see caption of Fig. \ref{fig:1}). As shown, most metabolites, both in \textit{E. coli} and \textit{M. pneumoniae}, present flux disparity values that cannot be explained by the null hypothesis, meaning that the local distribution of the fluxes of reactions associated to metabolites is significantly heterogeneous. We conclude then that the disparity filter will be able to efficiently extract the backbone with the most significant connections for both organisms, while preserving the characteristic features of metabolism as a complex network. Notice that significant fluxes are those with values much above the average expectation given by the null hypothesis and that, in absolute terms, they can be high or low (see Methods, section \ref{SB}).

The filter preserves a reaction in the backbone if the probability $\alpha_{ij}$ that its normalized flux $p_{ij}$ is compatible with the null hypothesis (\textit{p}-value) is smaller than a chosen significance threshold $\alpha$, which determines the filtering intensity. For each metabolite $i$, we compute the \textit{p}-value $\alpha_{ij}$  for each producing and consuming reaction $j$ and compare the obtained \textit{p}-value with the significance level $\alpha$. In this way, all the reaction with fluxes which are significant for the production or consumption of a metabolite can be selected, in contrast to the approach in \cite{Almaas:2004a} where only a single most significant flux was selected so that the obtained subgraphs presented an obvious linear structure. The disparity filter can be adjusted by tuning the critical threshold to observe how the metabolic networks of both \textit{E. coli} and \textit{M. pneumoniae} are reduced as we decrease $\alpha$ from 1 to 0, $\alpha=1$ meaning the complete genome-scale reconstruction. Notice that, after applying the filter with a specific value of $\alpha$, we recover a bipartite representation of the metabolic backbone. To avoid working with stoichiometrically non-balanced reactions, we transform the filtered bipartite representation into a one-mode projection of metabolites placing a directed link between two metabolites if there is a reaction whose flux is simultaneously significant for the consumption of one metabolite and for the production of the other \cite{Serrano:2009a} (see Fig. \ref{fig:1}c). 

Next, we compute the fractions of links $E$, metabolites $N$ and total flux $W$ remaining in the one-mode filtered networks as a function of the significance level $\alpha$. These magnitudes are normalized by dividing them by the corresponding values in the original network, $E_T$, $N_T$, and $W_T$. In Figs. \ref{fig:1}d and \ref{fig:1}e, we show $N/N_T$ vs $E/E_T$, and $W/W_T$ vs $E/E_T$ in the associated insets, for the one-mode projections of the filtered networks for both \textit{E. coli} and \textit{M. pneumoniae}. While the filter can reduce considerably the fraction of links, the corresponding fraction of metabolites is maintained at almost the original value. In addition, the total flux in the backbone only starts to drop appreciably after more than 50\% of the links are removed. We take the critical value $\alpha_c$ as the point where the fraction of metabolites starts to decay (see Figs. \ref{fig:1}d and e). This critical value can be seen as an optimal point which greatly reduces the number of links in the network preserving, at the same time, most nodes and therefore much biochemical and structural information as possible. The values are $\alpha_c=0.21$ for \textit{E. coli} and $\alpha_c=0.37$ for \textit{M. pneumoniae}. Notice that the value of $\alpha_c$ is lower for \textit{E. coli} due to its stronger heterogeneity in the local distribution of fluxes, which allows to reduce further the fraction of links while preserving all the nodes. The backbone of a metabolic network corresponds to the filtered one-mode projection graph using the critical value $\alpha_c$ as the \textit{p}-value threshold. The backbone unveils the pivotal pathways of fluxes in metabolism and at the same time preserves the connectivity structure, the heterogeneity of the degree distribution, and the high level of clustering typical of complex networks.

\subsection{Analysis of the large-scale structure and pathway composition of the metabolic backbones of \textit{E. coli} and \textit{M. pneumoniae}}
We filter the metabolic backbones of both \textit{E. coli} and \textit{M. pneumoniae} using the identified critical values for the significance level. The backbone of each organism retains all metabolites, 445 and 227 respectively. Next, we analyze their structure in terms of connectedness. Metabolic networks have been found to display typical large-scale connectivity patterns of directed complex networks, characterized by a bow-tie structure \cite{Broder:2000} (see Methods, section \ref{SC}), with most reactions in a interconnected core, named the strongly connected component (SCC), together with in (IN) and out (OUT) components formed mainly by nodes directly connected to the SCC component \cite{Ma:2003b, Serrano:2008}. This is the case also for the genome-scale reconstruction of both \textit{E. coli} and \textit{M. pneumoniae} used in this work, whose SCCs contain the largest part of metabolites and whose IN and OUT components are formed, respectively, by nutrients and waste metabolites.

\begin{figure*}
 \centering
 \includegraphics[width=0.95\textwidth]{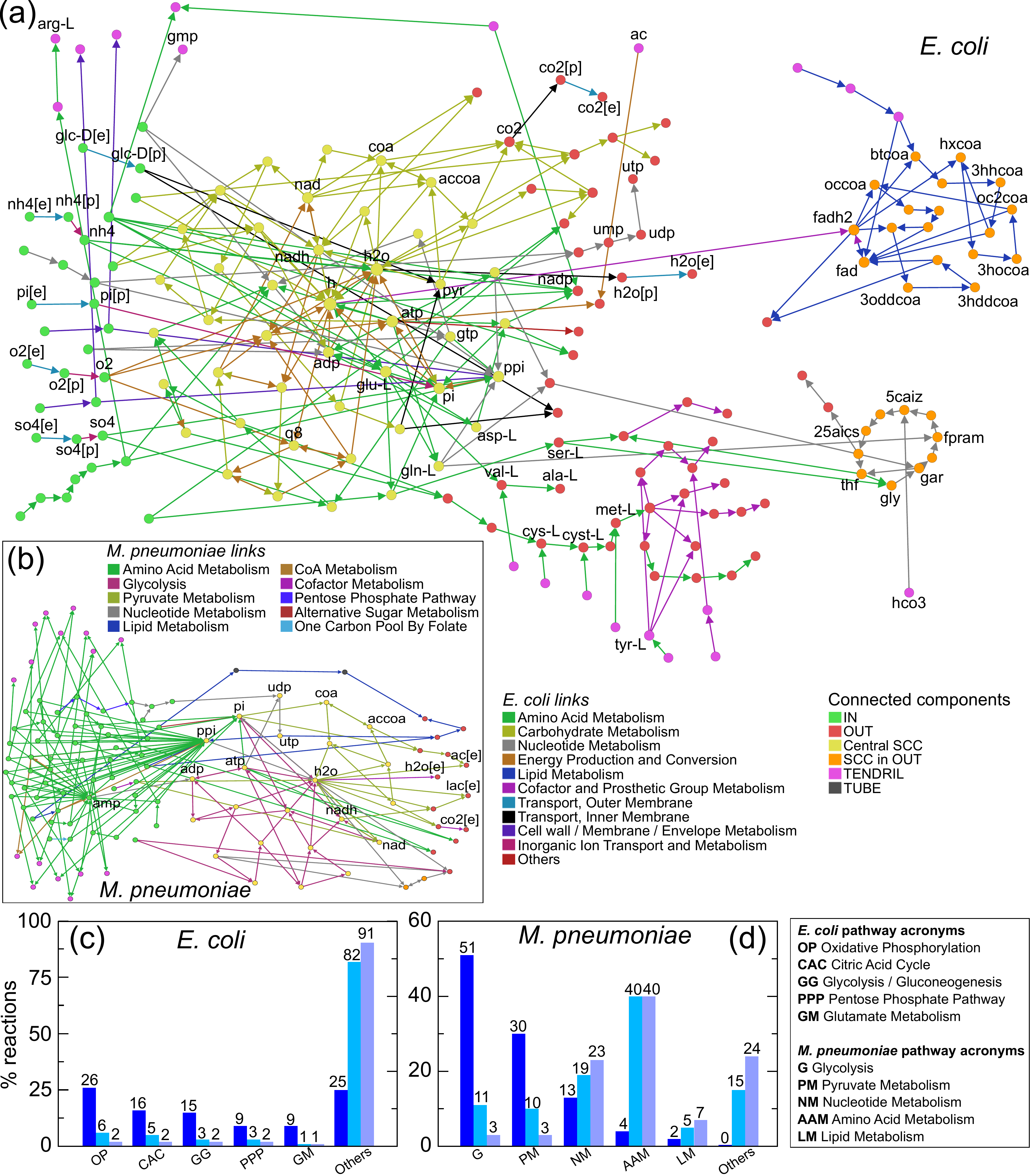}
 \caption{\textbf{Structure of connected component in the backbones of \textit{E. coli} and \textit{M. pneumoniae} and corresponding pathways.} (a) Connected components in the metabolic backbone of \textit{E. coli}. The colors of the metabolites depend on the component each node belongs to. The color of the links, and its association given in the legend, depends on the functional categories given in Ref.~\cite{Orth:2011a}, where each category contains pathways that realize similar tasks. (b) Connected components in the metabolic backbone of \textit{M. pneumoniae}. The color of the metabolites denote again the component each node belongs to. The color of the links, and its association given in the legend, depends on the pathway each reaction belongs to. (c) Percentage of links in pathways for the largest SCC (first columns, dark blue), for the backbone as a whole (second columns, light blue), and for the non-filtered metabolic network disregarding zero-flux reactions (third columns, violet) of \textit{E. coli}. (d) The same for \textit{M. pneumoniae}.}
 \label{fig:2}
\end{figure*}

Metabolites in the backbone of \textit{E. coli} are arranged in a large connected component of 178 metabolites and 51 disconnected small components. Three different SCCs can be identified in the largest connected component of the backbone, each one with 25\%, 10\%, and 6\% of the metabolites (see Fig. \ref{fig:2}a). The two smallest SCCs are in the OUT component of the largest SCC. Central compounds of metabolism are identified in these SCCs: protons, water, adenosine triphosphate (ATP), L-glutamate, phosphate, nicotinamide adenine dinucleotide (NAD$^+$), diphosphate, adenosine diphosphate (ADP) and flavin adenine dinucleotide (FAD$^+$). These metabolites are highly-connected even in the backbone, highlighting the ability of the disparity filter to preserve the same structural features of the complete metabolic network while in a reduced version. However, a comparison of the hubs in the backbone with the hubs in the non-filtered metabolic network, \textit{i.e.}, considering $\alpha=1$, shows some deviations. The most important difference happens for L-glutamate. It occupies the 4th position in the ranking of the most connected metabolites, whereas it is located in the 12th position in the non-filtered network. Besides, the fraction of reactions producing L-glutamate in relation to the total number of reactions in which this metabolite participates is 0.27 in the backbone vs 0.65 in the non-filtered network. Regarding other metabolites, which in this case cannot be considered as a hub due to its smaller degree, we find that FAD$^+$ ranks 12th in the backbone, while its position for the non-filtered case is 22nd. Recall that the number of metabolites in the backbone and in the non-filtered network is the same. Another interesting case concerns the metabolite glucose 6-phosphate. It contains no outgoing connections in the backbone, whereas its outgoing degree in the non-filtered network is four.

Since links in the metabolic backbone denote reactions, it is interesting to assess the composition of the backbone of \textit{E. coli} in terms of pathways. First, we start by computing the percentages of pathways in the original metabolic network. We find that the five pathways with more reactions are Transport, Inner Membrane (17\%), Cofactor and Prosthetic Group Biosynthesis (13\%), Glycerophospholipid Metabolism (12\%), Alternate Carbon Metabolism (8\%), and Cell Envelope Biosynthesis (8\%). The same analysis in the metabolic backbone gives a different composition, with pathways such as Oxidative Phosphorylation and the Citric Acid Cycle gaining weight, as shown in Fig. \ref{fig:2}c. This is in fact due to the pathway participation in the largest SCC, where the major contribution comes from these two pathways along with Glycolysis/Gluconeogenesis and Pentose Phosphate Pathway. It has been found that Glycolysis and Pentose Phosphate Pathway can take place without the need of enzymes \cite{Keller:2014a}. Concerning the Citric Acid Cycle, it is an ancient pathway that has evolved in order to achieve maximum ATP efficiency \cite{Melendez-Hevia:1996a} by being coupled to Oxidative Phosporylation and Glycolysis. In addition, this coupling helps the organism to decrease their quantity of reactive oxygen species by modulation of their participating metabolites \cite{Mailloux:2007a}, conforming in this way one of the central pillars of carbon metabolism and energy production. Another pathway significantly present in the largest SCC is Glutamate Metabolism. Glutamate has been reported to be one of the oldest amino acids used in the earliest stages of life \cite{Fell:2000a}. The second largest SCC contains links that belong mainly to Membrane Lipid Metabolism (97\%) and Cofactor and Prosthetic Group Biosynthesis (3\%). Membrane Lipid Metabolism supplies the necessary lipids to generate the cell membrane needing the participation of the cofactor FAD$^+$/FADH$_2$. It has been shown that the pathways involved in lipid metabolism exhibit differences between different lineages in organisms \cite{Suen:2012a}, whereas pathways related to central metabolism are more conserved \cite{Suen:2012a}. Finally, pathways related to the smallest SCC are Purine and Pyrimidine Biosynthesis (91\%). Purines and pyrimidines serve as activated precursors of RNA and DNA, glycogen, etc. \cite{Evans:2004a,Powner:2009a}. It has been found that the synthesis of purines and pyrimidines was the first pathway involving enzyme-based metabolism \cite{Caetano-Anolles:2007a}. Interestingly, the other contribution to this SCC is Glycine and Serine Metabolism. Glycine is a precursor of purines and pyrimidines. 

When considering $\alpha$ values smaller than the critical one, implying that the filter is more restrictive, we observe that the smallest SCCs disappears from the more stringent sub-backbone. More precisely, it happens for a value of $\alpha=0.19$. Decreasing even more the significance level to $\alpha=0.15$, the second largest SCC containing reactions in the Purine and Pyrimidine Biosynthesis pathway still retains 30\% of the nodes in the backbone, whereas the largest SCC still contains a 86\%. At a value of $\alpha=0.14$, the second SCC finally disappears and the sub-backbone only remains a single SCC, still preserving 82\% of the nodes in the backbone. Hence, the SCC containing links belonging to pathways related to energy metabolism shows a large resistance to get fragmented, even though the filter becomes progressively more and more restrictive, which points to increased levels of local flux heterogeneities and so of flux dependencies for the consumption and production of metabolites.

We perform the same analysis in \textit{M. pneumoniae}, a simpler organism which has been proposed as a bacterial model due to its simplicity and its reduced genome \cite{Yus:2009a,Wodke:2013a}. The non-filtered metabolic network is dominated by reactions in Amino Acid and Nucleotide Metabolism, consistent with the composition of the backbone. The connected component of the backbone is shown in Fig. \ref{fig:2}b. It contains two SCCs, one of them with only two metabolites (see Fig. \ref{fig:2}b). The relevant SCC contains 21\% of the metabolites in the connected component, and most of its links are related also with energy metabolism, as in \textit{E. coli}. The dominant pathways in the SCC are Glycolysis and Pyruvate Metabolism, see Fig. \ref{fig:2}d where this composition is compared with that of the backbone as a whole. Both Glycolysis and Pyruvate Metabolism, which dominate the composition of the core of the backbone, are pathways that were present in the earliest stages of life \cite{Tadege:1999a}, when no oxygen was present in the early atmosphere. 
\\

\subsection{The metabolic backbone of \textit{E. coli} encodes its adaptation capabilities}
In the previous section, we analyzed the pathway composition of the metabolic backbone of \textit{E. coli} in glucose minimal medium and found that pathways in the central core are related to energy metabolism. The analysis of the backbone also revealed that the synthesis of nucleotides and the metabolism of lipids form smaller cores which rely critically on energy metabolism; but not conversely. In this section, we study how changes in the environment modify the backbones of \textit{E. coli} in relation to the one obtained in glucose minimal medium, which exposes adaptation capabilities.

\begin{figure*}
 \centering
 \includegraphics[width=0.8\textwidth]{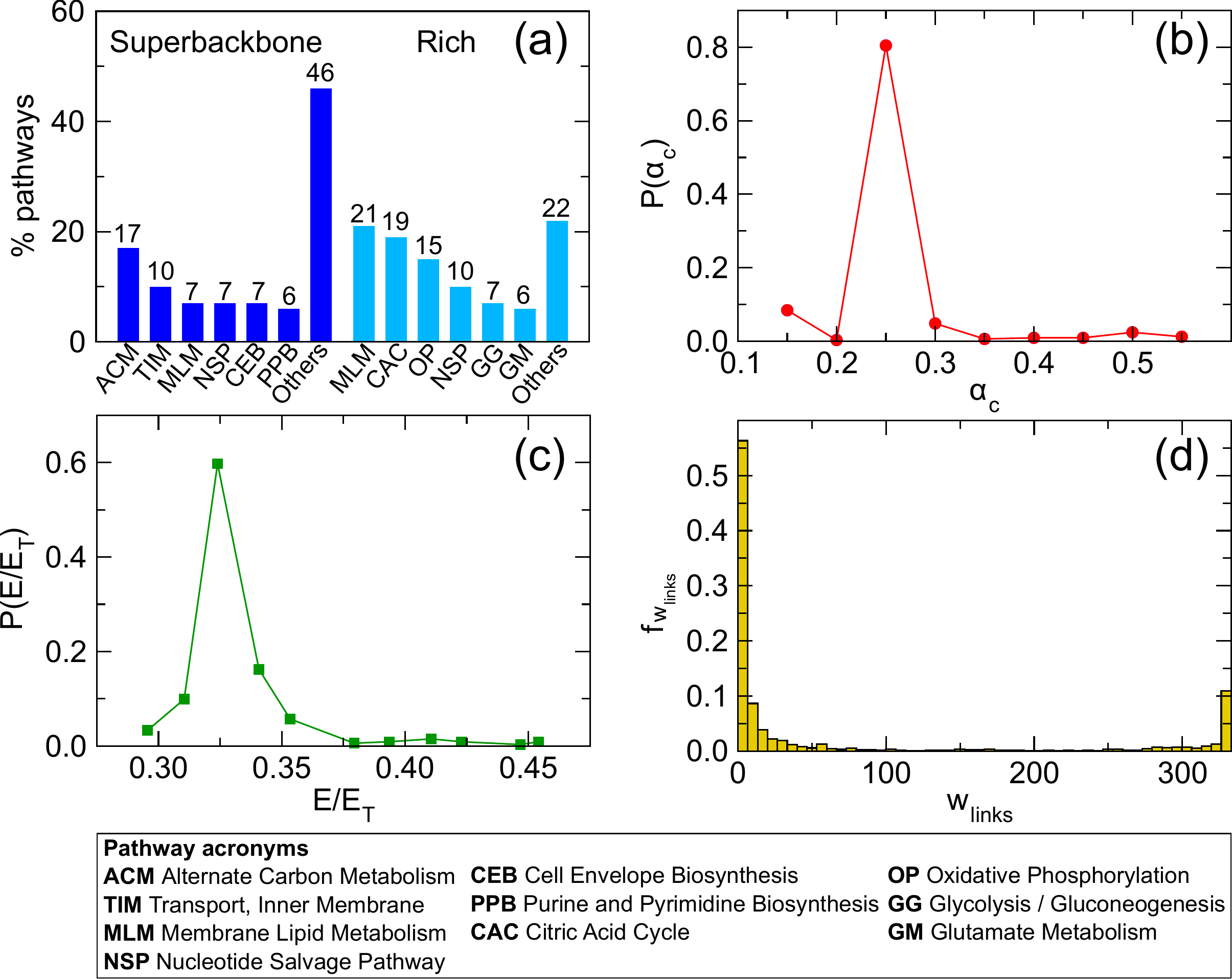}
 \caption{\textbf{Dependence of the backbone of \textit{E. coli} with the composition of the environment.} (a) Histogram of the fraction of links belonging to each pathway (x axis) for  the largest SCC of the backbone in the 333 minimal media (blue) and in the rich medium (sky blue). (b) Probability distribution function of $\alpha_c$ for all minimal media. (c) Probability distribution function of the fraction of links in the metabolic backbones for all minimal media. (d) Histogram of relative frequency of weights of links in the metabolic superbackbone, where the weight of the links counts the number of media in which the corresponding metabolic backbone contains the link.}
 \label{fig:3}
\end{figure*}

First, we calculate the FBA fluxes that maximize the growth rate of \textit{E. coli} in the rich medium Luria-Bertani (LB) Broth \cite{Sezonov:2007a,Guell:2014b} (see Methods, section \ref{SD}). Afterwards, we apply the disparity filter to extract the metabolic backbone in this new environment, that is obtained for a significance critical threshold $\alpha_c=0.4$. This value is noticeably larger than $\alpha_c=0.21$ identified for the glucose minimal medium. Interestingly, this rich medium activates 400 reactions, 11 less than in glucose minimal medium. Of them, 279 are active in both media, of which 247 have a larger flux in LB Broth. An analysis of the connected components in the metabolic backbone of \textit{E. coli} in rich medium is also performed. We find that it contains a large connected component with 188 metabolites and 60 small disconnected components. The connected component contains also three SCCs. However, two of them are tiny with only two metabolites, whereas the largest one encloses 34\% of the metabolites in the connected component. Interestingly, the pathway contributing more reactions to this large SCC is Membrane Lipid Metabolism (see Fig. \ref{fig:3}a). This fact is in accordance with Ref. \cite{Tao:1999a}, where the authors found that the expression of the genes which synthesize fatty acids was generally elevated in rich medium. Another important difference is the relative loss of prominence of Oxidative Phosphorylation and the Pentose Phosphate Pathway.  This might seem surprising since the Pentose Phosphate Pathway is typically the main source of nicotinamide adenine dinucleotide phosphate (NADPH). However, in rich medium the functionally significant production of this metabolite takes place in the Citric Acid Cycle pathway \cite{Sauer:2004a}. This also evidences the importance of the Citric Acid Cycle to produce NADPH, and so its importance in the synthesis of membrane lipids. Nevertheless, links associated to both Oxidative Phosphorylation and Pentose Phosphate Pathway are also present in the backbone, located outside the SCCs. 

Next, we consider the set of minimal media given in Ref. \cite{Orth:2011a} (see Methods, section \ref{SD}) where different carbon, nitrogen, phosphorus and sulfur sources are alternated. For each minimal medium, we scan for $\alpha_c$ as in Figs. \ref{fig:1}d and  \ref{fig:1}e. In Fig. \ref{fig:3}b and \ref{fig:3}c we plot, respectively, the probability distribution function of the collection of $\alpha_c$ values and of the fraction of links remaining in the metabolic backbone for all media. We find that these magnitudes present a characteristic value, meaning that the flux structure is very similar across media in spite of the difference in the composition of nutrients. The presence of these characteristic value of $\alpha_c$ and the stability of the retained fraction of links in the metabolic backbone in the different media motivate us to combine all of them into a single merged metabolic backbone, obtaining a network which we call superbackbone. The links in this superbackbone correspond to reactions that passed the filter in any of the external media considered, and are annotated with a weight that corresponds to the number of media in which the corresponding metabolic backbone contains the link. The histogram of the distribution of these weights is shown in Fig. \ref{fig:3}d, characterized by a clear bimodal behavior. One peak corresponds to links being common to all media, and the other corresponds to the most common situation of links specific to a few media.

An analysis of connectedness shows that this superbackbone contains a large connected component and 11 disconnected components. The connected component with 1090 metabolites is composed by a large SCC containing 43\% of its metabolites, in addition to three small SCCs containing only two nodes each. A pathway composition analysis in the large SCC indicates that, again, we obtain significantly different results from the glucose minimal medium (see Fig. \ref{fig:2}c and Fig. \ref{fig:3}a). The most prominent pathway is Alternate Carbon Metabolism, in agreement with Ref. \cite{Lourenco:2011a}, where the authors found that Alternate Carbon Metabolism is related to genes whose expression depends on external stimuli, particularly on alteration of carbon sources. It is also in agreement with results in Ref. \cite{Monk:2013a}, where the authors hypothesize that Alternate Carbon Metabolism can adapt to different nutritional environments, and also with results in Ref. \cite{Serrano:2012a}, where Alternate Carbon Metabolism is found to be an important intermediate pathway in the network of pathways. The second most abundant pathway corresponds to Transport, Inner Membrane, which again is in agreement with Ref. \cite{Lourenco:2011a} and Ref. \cite{Serrano:2012a}. This pathway is in charge of the transport of metabolites between periplasm and cytosol. Finally, if we retain links present at least in 25\% of the minimal media, the network fragments into 40 components with the largest one containing five SCCs, which indicates that links with small weight, \textit{i.e.} links specific for a few media, have an important role in order to prevent the fragmentation of the superbackbone.

\section{Discussion}
Identifying high-flux routes in metabolic networks has been useful in order to, for example, identify principal chains of metabolic transformations \cite{Almaas:2004a,Bourqui:2007a,Faust:2010a}. In this work, we go beyond the mere identification of high-flux routes with metabolic pathways. We are able to identify pivotal pathways containing both high and low fluxes which are central for the maximization of the growth rate and, on the other hand, to identify pathways more responsive to changes in the environment. To perform the analysis, we used FBA and a filtering tool which needs no \textit{a priori} assumptions for the connectivity of the filtered network, but which produces a backbone as a reduced version of the metabolic network. The backbone is globally connected and retains the characteristic complex features of the metabolic network as a whole. It contains all functionally significant connections given a set of external nutrients, recovering both intra- and inter-pathway connections which mark the most significant routes in metabolism formed by the most important interdependencies for the production and consumption of metabolites. As an explanation for the the strong dependence between between both intra- and inter-pathway connections present in the backbone, one could say that the overall performance of a cell is not only improved due to a better and more efficient utilization of the available resources but depends critically on cross interactions. This fact reinforces the idea that pathways are not isolated identities performing their tasks independently of one another \cite{Serrano:2012a}. 

As stated in Ref. \cite{Basler:2011a}, properties that originate from evolutionary pressure should not be observed in random networks. In our investigations, the effect of evolutionary pressure is understood to favor the maximization of the growth of the organism \cite{Ibarra:2002a,Blank:2005a,Llaneras:2008a}, in accordance to FBA and its assumptions. It is important to note that the assumed objective function used to compute FBA can have a direct effect on the diversity of fluxes in the metabolic network and so in the proportions of pathways in the filtered backbone and its cores. The analysis of the metabolic backbone of \textit{E. coli} and \textit{M. pneumoniae} in glucose minimal medium shows that, for both organisms, the SCCs is mainly composed by reactions that belong to ancient pathways that were already present at the first stages of life. A very different composition is obtained when considering the backbone as a whole (see Fig. \ref{fig:2}c and Fig. \ref{fig:2}d) and even more when comparing with the non-filtered metabolic network. In the backbone of \textit{E. coli}, each of the three SCC has a different and definite metabolic function. The largest SCC contains pathways related to energy metabolism, as for \textit{M. pneumoniae}, and its composition is dominated by the Citric Acid Cycle, Oxidative Phosphorylation, and Glucolysis/Gluconegenesis, in accordance with the fact that the Citric Acid Cycle has evolved towards an optimal chemical design \cite{Melendez-Hevia:1996a} by coupling to Oxidative Phosporylation and Glycolysis \cite{Ebenhoh:2001a}. A second SCCs corresponds to the metabolism of lipids, the most important constituents that compose the cell membrane. The smaller SCC is responsible for the synthesis of purines and pyrimidines, vital for DNA / RNA synthesis. Two findings relating the two small SCCs deserve also special attention. Firstly, the two small SCCs are located in the OUT component of the large SCC. Secondly, as the filter becomes more restrictive, the small SCCs fragment, while the large SCC still maintains a large part of links and nodes. These features could be explained in terms of functional requirements. On the one side, the smaller cores need chemical energy to perform their tasks and, on the other side, they need also basic building blocks. These needs are covered by the large SCC, which suggests that the smaller SCCs were added to the OUT component of the largest SCC in later steps of evolution. An analysis of the connectivity of the hubs reveals that L-glutamate acquires in the backbone a functional relevance as a reactant which emphasizes its importance as the amino group donor for nearly all nitrogen-containing metabolites of the cell, in contrast to its role in the non-filtered network, where it could be classified rather as a product since in minimal medium it must be synthesized. The case of glucose 6-phosphate is also very interesting in the sense that it presents no outgoing connections in the backbone, but it is known that this metabolite has many metabolic fates. Hence, we conclude that, in glucose minimal medium, no route consuming this metabolite can be considered more significant than the others, being thus all of them equally significant. Therefore, glucose 6-phosphate cannot be considered as a functional precursor but a functional product, a fact confirmed due to its presence in the OUT component of the backbone. A simpler organism, \textit{M. pneumoniae}, has no other relevant SCCs apart from energy metabolism, as a result of its parasitism, which has led to the loss of many metabolic functions \cite{Wodke:2013a}. More precisely, in \textit{M. pneumoniae} the Citric Acid Cycle and Oxidative Phosphorylation do not take place \cite{Manolukas:1988a,Wodke:2013a}, meaning that it must rely on organic acid fermentation to obtain energy. Moreover, changes in the growth rate greatly affect the fluxes through Glycolysis and Pyruvate Metabolism \cite{Wodke:2013a}.

The study of the dependence on the environment of the \textit{E. coli} metabolic backbone allows us to identify adaptation capabilities. Regarding rich medium, we observe that the critical value of $\alpha$ is substantially different than the one in glucose minimal medium, suggesting that this enriched medium modifies significantly the flux structure compared to the glucose minimal medium. The bacterium in rich medium displays less active reactions than in glucose minimal medium since, in minimal medium, many reactions must be active in order to synthesize biosynthetic precursors that in the rich medium can be obtained from the environment, in agreement with Ref. \cite{Tao:1999a}. The pathway called Membrane Lipid Metabolism achieves a high relevance, being the most abundant pathway in the largest core of the rich medium metabolic backbone. This may happen because the proliferation response of \textit{E. coli} to the rich medium increases lipid biosynthesis, since lipids are not present as nutrients even in rich media due to its water insolubility. In contrast, biosynthesis of other components, specially amino acids, is obviously reduced. To satisfy the high lipid demand to generate new cells \cite{Tao:1999a}, fast-growing cells must synthesize membrane lipid components more rapidly. This leads to an over-expression of the genes related to membrane lipids which, in terms of metabolism, is observed in terms of a high relevancy of the Membrane Lipid Metabolism pathway. Another feature that we find is the loss of prominence of the Pentose Phosphate Pathway. Although this may suggest to a deficiency of the metabolite NADPH, we find that, in the rich medium backbone, the functionally significant production of this metabolite takes place in the Citric Acid Cycle pathway \cite{Sauer:2004a}. This evidences the importance of Citric Acid Cycle to produce NADPH, which is tightly related to the synthesis of membrane lipids. An alternative explanation for the increase of the appearance of the Citric Acid Cycle in the backbone is that more metabolites need to be catabolized in rich medium, and the TCA cycle is the primary site for catabolism of many non-sugar substrates through anaplerotic reactions. The analysis of the adaptation of \textit{E. coli} to 333 different minimal media shows that the fraction of links in the backbone is practically independent on the composition of the nutrients present in these environments (see Fig. \ref{fig:3}c). This permits the construction of a merged backbone that comprises all the links in the backbone in each different media and in which each contributing backbone has uniform representation in fraction of links. This leads to the identification of pathways whose associated reactions are more sensitive to changes in the environment, unveiling Alternate Carbon Metabolism as the pathway with more capabilities to respond to external stimuli, in accordance with existing works \cite{Lourenco:2011a,Monk:2013a}. This finding could also be seen as a successful positive control for the methodology applied here.

The use of filtering methods usually implies a drastic reduction of the complexity of metabolic maps, which weakens the validity of potentially inferred conclusions. In contrast, the application of the disparity filter, based on a flux significance analysis to produce metabolic backbones, enables to reduce the system while maintaining all significant interactions according to $\alpha_c$ and so it becomes a useful tool to unveil sound biological information. Notice that these reduced versions must be seen as a map of the most significant connections, even though other metabolic reactions must be present to achieve viability and to satisfy the physico-chemical laws governing metabolic networks depending on nutritional conditions and other stresses applied to the cell. Our investigations of \textit{E. coli} and \textit{M. pneumoniae} revealed metabolic backbones mainly composed of a core of reactions belonging to ancient pathways that still retain at present a central role in the evolved metabolism. Besides, in \textit{E. coli}, the analysis of the core reveals a dominant direction with the synthesis of purines and pyrimidines and the metabolism of lipids ensuing after energy metabolism. At the same time, our approach can be appropriate in applications exploiting its capability to recognize pathways and particular reactions more sensitive to environmental changes, which makes it potentially useful in biotechnology and biomedicine.

\section{Methods}\label{meth}
\subsection{Flux Balance Analysis}\label{SA}
Flux Balance Analysis (FBA) \cite{Orth:2010a} is a technique which allows to compute metabolic fluxes without the need of kinetic parameters, just by using constrained-optimization. FBA proceeds by writing the stoichiometric matrix $S$ of the whole network and multiplying it by the vector of fluxes $\nu$. This stoichiometric matrix contains the stoichiometric coefficients of each metabolite in each reaction of the network. This product is then equal to the vector of the time variation of the concentrations $\dot{c}=S\cdot \nu$. Steady-state is assumed, thus $S \cdot \nu=0$. Since in general metabolic networks contain more reactions than metabolites, we have an underdetermined system of equations. Hence, a biological objective function must be defined in order to have a biologically meaningful solution. In this work, the chosen objective function is the growth rate of the organism, which means that FBA finds the solution that optimizes the growth of the organism, which is equivalent to maximize biomass formation. Reversibility of reactions are also added in order to constrain the solutions. Since we have a linear system of equations with linear constraints, Linear Programming is used in order to compute a flux solution in a small amount of time (of the order of 1 s), which implies a computationally cheap method.

We implement FBA using GNU Linear Programming Kit (GLPK). FBA calculations are performed on the \textit{i}JO1366 version of \textit{E. coli} K-12 MG1655 \cite{Orth:2011a}, which contains 1805 metabolites and 2583 reactions, and on the \textit{i}JW145 version of \textit{M. pneumoniae} \cite{Wodke:2013a}, which contains 266 metabolites and 306 reactions. In both organisms, the versions include the growth reaction, auxiliary reactions such as exchange or sink reactions, and all cellular compartments are taken into account. We model these metabolic networks as bipartite graphs with two kind of nodes, metabolites and reactions, and with links containing directionality, which leads to incoming, outgoing, and bidirectional links. For \textit{E. coli}, FBA calculations are performed in a glucose minimal medium with a maximum uptake of glucose limited to 10 mmol gDW$^{-1}$ h$^{-1}$ \cite{Orth:2011a}, whereas for \textit{M. pneumoniae}, FBA computations are carried in a defined medium with a maximum glucose uptake of 7.37 mmol gDW$^{-1}$ h$^{-1}$ and a supply of D-ribose to simulate the availability of ribosylated bases \cite{Wodke:2013a}.

\subsection{Disparity filter on metabolic networks}\label{SB}
The disparity filter \cite{Serrano:2009a} takes advantage of the local heterogeneity present in the fluxes of reactions associated to a given metabolite. The filter is able to retain those fluxes which are significant with respect to a null hypothesis. Notice that, since we work with directed metabolic networks, we have three kinds of links: incoming, outgoing and bidirectional links. The latter are decoupled into incoming and outgoing links, leading to a network containing only incoming and outgoing links. We treat incoming and outgoing connections separately. The filtering method starts by normalizing the $k$ fluxes $\nu_{ij}$ producing or consuming a metabolite $i$, $p_{ij}=\frac{\nu_{ij}}{\sum_{j=1,k}\nu_{ij}}$. The key point is that a few links carry a significant value of $p_{ij}$. We characterize the disparity in the local distribution of fluxes around $i$ with the measure \cite{HHIHerfindahl:1959,HHIHirschman:1964}

\begin{equation*}
 \Upsilon_i(k) \equiv k \sum_j p_{ij}^2.
\end{equation*}

Under perfect homogeneity, when all the links share the same amount of the  flux producing or consuming the metabolite, $\Upsilon_i(k)=1$ independently of $k$, whereas for perfect heterogeneity, when one of the links carries all the flux, $\Upsilon_i(k)=k$. Usually, an intermediate behavior is observed in real systems.

To assess the  relevance of the difference fluxes associated to a metabolite, a null hypothesis is used which provides the expectation of the disparity measure. The null hypothesis consists on  assuming that the intensity of the different fluxes associated to a metabolite is consistent with a random assignment coming from a uniform distribution. Incoming and outgoing links are treated independently. The filter then proceeds by identifying which fluxes must be preserved. To do this, we compute the probability $\alpha_{ij}$ that a normalized flux $p_{ij}$ is non-compatible with the null hypothesis. This probability is a \textit{p}-value which is compared with a significance level $\alpha$, and thus links that carry fluxes with a probability $\alpha_{ij}<\alpha$ can be considered non-consistent with the null model and so significant for the metabolite. The probability $\alpha_{ij}$ is computed with the expression (see Ref. \cite{Serrano:2009a}).

\begin{align*}
 \alpha^{in}_{ij}&=(1-p^{in}_{ij})^{k^{in}-1} \\
 \alpha^{out}_{ij}&=(1-p^{out}_{ij})^{k^{out}-1}
\end{align*}

Once the significant fluxes have been selected, we construct the backbone by placing a directed link between two metabolites if there is a reaction whose flux is simultaneously significant for the consumption of one metabolite and for the production of the other. The filter cannot decide on nodes with just one connection. According to our maximization of nodes minimization of links principle, we use the prescription to preserve the reaction associated to metabolites with only one incoming or outgoing connection.

\subsection{Connected components}\label{SC}
A connected component of an undirected network is a subset of the network in which any two nodes are connected by at least one path. Nodes in a connected component do not share connections with nodes belonging to a different connected component \cite{Newman:2010}.

Directedness in network connections introduces a rich substructure in the connected components. Inside the connected component of a directed network, the so-called bow-tie structure emerges \cite{Broder:2000}. Bow-tie structures are formed by a strongly connected component (SCC), IN and OUT components, tubes and tendrils. A SCC is a subset of the connected component where nodes are reachable from any other by a directed path. The IN component contains nodes that can access the SCC but not vice versa. The OUT component is formed by nodes that can be reached from the SCC but that cannot return there. A tube is a sequence of nodes that connect the IN and the OUT component without going through the SCC. Tendrils are composed by nodes that have no access to the SCC and are not reachable from it.

\subsection{Construction of environments in \textit{E. coli}}\label{SD}
\subsubsection{Luria-Bertani Broth}
We consider a rich medium called Luria-Bertani Broth. This nutritionally rich medium contains the set of compounds defining the minimal medium \cite{Orth:2011a}, \textit{i.e.}, a set of minerals salts and four metabolites representing carbon, nitrogen, phosphorus and sulfur sources, in addition to the following compounds: amino acids, purines and pyrimidines, biotin, pyridoxine, thiamin, and the nucleotide nicotinamide mononucleotide (see Ref. \cite{Guell:2014b} for specific details).

\subsubsection{Minimal media}
We use the different minimal media defined in Ref. \cite{Orth:2011a}. More precisely, these media contain a set of minerals salts and four extra metabolites representing carbon, nitrogen, phosphorus and sulfur sources \cite{Orth:2011a}. To determine FBA solutions in specific media we change the carbon source while we fix the sources of nitrogen, phosphorus and sulfur to the standard metabolites in each class, which are ammonia, phosphate, and sulfate respectively. In this way, each carbon source determines a different minimal medium. In the same way, other minimal media are constructed using the same procedure of changing the nitrogen, phosphorus or sulfur source while keeping the standard metabolite unchanged for the rest of categories; note that, in these cases, the standard carbon source is fixed to glucose. 555 media can be constructed using this procedure, 333 of them allowing growth after computing the FBA solution in each media.

\section{Acknowledgments}
We acknowledge support from a James S. McDonnell Foundation 21st Century Science Initiative in Studying Complex Systems Scholar Award; MINECO Projects No. FIS2013-47282-C02-01 and No. FIS2013-41144-P; and the {\it Generalitat de Catalunya} grants No. 2014SGR608 and No. 2014SGR597. O.~G. acknowledges support from a FPU grant funded by MINECO.


%

\end{document}